\documentclass[12pt,a4paper]{article}

\pdfoutput=1

\usepackage[english]{babel}
\usepackage{natbib}
\usepackage{authblk}
\usepackage{amssymb,amsfonts,textcomp}
\usepackage{amsmath}
\usepackage{bm}
\usepackage{bbm}
\usepackage{xspace}
\usepackage{hyperref}

\usepackage[pdftex]{graphicx}

\graphicspath{{./pics/}} 

\begin{document}
\newcommand{\ie}{\emph{i.e.}\@\xspace}
\newcommand{\eg}{\emph{e.g.}\@\xspace}
\newcommand{\E}{\mathbb{E}}
\newcommand{\V}{\mathbb{V}}
\newcommand{\1}{\mathbbm{1}}
\newcommand{\bG}{\bold{G}}
\newcommand{\bY}{\bold{Y}}
\newcommand{\by}{\bm{y}}
\newcommand{\bx}{\bm{x}}
\newcommand{\bw}{\bm{w}}
\newcommand{\bI}{\bold{I}}
\newcommand{\R}{\textsf{R}\xspace}

\newcommand{\bbeta}{\bm{\beta}}
\newcommand{\bepsilon}{\bm{\epsilon}}

\title{Inference for feature selection using the Lasso with high-dimensional data}
\author[1]{Kasper Brink-Jensen}
\author[2]{Claus Thorn Ekstr\o m}
\affil[1]{Department of Mathematical Sciences, University of Copenhagen}
\affil[2]{Department of Biostatistics, University of Copenhagen}

\maketitle

\begin{abstract}
\textbf{Motivation:}
  Penalized regression models such as the Lasso have proved useful for
  variable selection in many fields --- especially for situations with
  high-dimensional data where the numbers of predictors far exceeds
  the number of observations. These methods identify and rank
  variables of importance but do not generally provide any inference
  of the selected variables. Thus, the variables selected might be the
  ``most important'' but need not be significant. We propose a
  significance test for the selection found by the Lasso.

\textbf{Results:}
We introduce a procedure that computes inference and $p$-values for
features chosen by the Lasso. This method rephrazes the null
hypothesis and uses a randomization approach which ensures that the
error rate is controlled even for small samples.  We demonstrate the
ability of the algorithm to compute $p$-values of the expected
magnitude with simulated data using a multitude of scenarios that
involve various effects strengths and correlation between
predictors. The algorithm is also applied to a prostate cancer dataset
that has been analyzed in recent papers on the subject.  The proposed
method is found to provide a powerful way to make inference for
feature selection even for small samples and when the number of
predictors are several orders of magnitude larger than the number of
observations.

\textbf{Availability:}
The algorithm is implemented in the MESS package in \R and is freely
available.

\textbf{Contact:} \href{ekstrom@sund.ku.dk}{ekstrom@sund.ku.dk}
\end{abstract}

\section{Introduction}
Molecular technologies have reached a state where it is possible to
quickly and cheaply measure the status of millions of nucleotides,
mRNAs, genes, proteins, or metabolites simultaneously and repeatedly
over time. The availability of these massive data ideally allows for
complex and detailed modeling of the underlying biological system but
they also pose a serious potential multiple testing problem because
the number of covariates are typically orders of magnitude larger than
the number of observations. Recently, investigators are starting to
combine several of these high-dimensional dataset which has increased
the demand for analysis methods that accommodates these vast data
\citep{nie2006integrated,plosone2013,su:etal:2011,kamb:etal:2011}.

As data sets increase in size, methods that promote sparse results
have become more important and popular. \citet{hastie2005elements} and
\citet{hesterberg2008least} provide an overview of the development of
these methods of which the Lasso (least absolute shrinkage and
selection operator) is the most widely used
\citep{tibshirani1996regression}.  While all these estimators enforce
sparsity (since only a fraction of the variables are believed to be
influential) and perform variable selection, they lack the inference
from traditional statistics, such as $p$-values and confidence
intervals. As a consequence, these models will always identify a set
of the most important predictors even if none or only some of them are
significant.

To address this issue several recent papers propose techniques that
help asses the importance of the selected predictors.
\citet{Lockhart:2013fk} developed a method that use a covariance test
statistic approach to compute $p$-values for parameters obtained from
a Lasso regression. With this method $p$-values can be calculated for
all predictors in a data set if there are more samples than
predictors, $p<n$. In the case of $p>n$ it is not possible to use the
covariance test without specifying an estimate of $\sigma^2$, the
error standard deviation.  \citet{meinshausen2009p} presents another
approach where the data is split into two groups. The Lasso is applied
to one group, after which the variables selected by the Lasso are used
as predictors to obtain $p$-values from an ordinary least squares
(OLS) regression on the other group.  \citet{Wu2009}
suggest to first use the Lasso to select a set of relevant variables
which are then fitted to a non-penalized model where from which the
$p$-values can be calculated.

In applied sciences we find often that variable selection is used for
screening to determine which predictors such as genes to focus on, in
further (possibly costly) analyses. A useful tool in this setting
would be a model that allowed us to perform variable selection and
assign a $p$-value to evaluate the evidence of the most important
predictors found from the variable selection procedure. We wish to
propose an approach for variable selection and inference for the Lasso
that works directly when $p>n$ rather than choosing variables with
Lasso and transferring these to an OLS domain for inference. Our
approach uses randomization/resampling to infer the
significance of the $k$th chosen predictors from a Lasso model. Thus,
our focus is on using the Lasso to not only generate new (biological)
hypotheses but also to evaluate the evidence for those hypotheses.

The paper is structured as follows: in the next section we present a
method to both identify features and for computing $p$-values of those
features obtained from a Lasso regression model. Potential problems
and possible extensions are also discussed.  The simulation section
presents results from a series of simulation studies that show the
power of the proposed method under various conditions and for
situations both with and without correlated predictors. In addition we
show how well the method identifies the most important of the
predictors and apply the proposed method to a dataset on prostate
cancer. Finally, the proposed method is discussed. Note that while we
use the Lasso in the following the proposed procedure can essentially
be applied in combination with any sparse feature selection approach
(e.g., elastic net, ridge regression).

%\begin{methods}
\section{Methods}

We consider the situation where we have a quantitative response
vector, $\by \in \mathbb{R}^{n}$ and a set of $p$ quantitative
predictors $x_{i1}, \ldots, x_{ip}$ for each observation $i \in \{1,
\ldots, n \}$. The number of predictors can be much larger than the
number of observations, $p\gg n$ and we are interesting in identifying
predictors that are associated with the response vector $\by$.

Suppose that $\by = X\bbeta + \bepsilon$ where the noise is assumed to
be Gaussian, $\bepsilon\sim N(0, \sigma^2 \bI)$.  The Lasso  is a penalized version of
the multiple regression model where sparsity of the parameter vector
$\bbeta\in \mathbb{R}^{p}$ is achieved by adding a penalty term to
solve
  \begin{equation}
\hat{\bbeta}=\underset{\bbeta \in \mathbb{R}^p}{\operatorname{arg\,min}}
\left\{ \frac{1}{2}\|\by-X\bbeta\|^{2}_{2}+\lambda\|\bbeta\|_{1} \right\},
\label{eq:lasso}
\end{equation}
where $\lambda\ge0$ is the regularization parameter
\citep{tibshirani1996regression}. $\lambda$ 
controls the amount of shrinkage of $\hat{\beta}$ such that $\lambda
=\infty$ corresponds to setting all parameters to 0, while $\lambda=0$
corresponds to an ordinary least squares multiple regression
model with no restrictions on the parameter vector. 

The choice of $\lambda$ will be discussed below.

The Lasso can be used for variable selection for high-dimensional data
and produces a list of selected non-zero predictor variables.
\citet{Meinshausen2009} show that while the Lasso may not recover the
full sparsity pattern when $p\gg n$ and when the irrepresentable
condition is not fulfilled (e.g., if highly correlated predictor
variables are present) it still distinguishes between important
predictors (i.e., those with sufficiently large coefficients) and
those which are zero with high probability under relaxed conditions.

In many high-dimensional molecular genetic datasets it is reasonable
to assume that there is just a small number of genes that influence the
outcome while the vast majority of genes are irrelevant for the
outcome. This common setup is also relevant in other situations and it
forms the basis of the following.  When the relative number of truly
non-zero predictors is small then the results from
\citet{Meinshausen2009} also show that the Lasso selects the non-zero
parameters as well as ``not too many additional zero entries of
$\bbeta$''. Here, we seek to identify which of the selected predictors
that we believe to be non-zero. The predictors found by Lasso are
generally ranked by their importance, so that a high value for
$\lambda$ will include the most important predictors, decreasing
$\lambda$ will include other less important predictors
\citep{efron2004least, hastie2005elements, hesterberg2008least}.

We wish to determine which of the features identified by the Lasso
that are indeed true positives. Let $\beta_{(1)}, \ldots, \beta_{(k)}$
be the ordered absolute coefficients of the $k$ non-zero predictors that are
selected by the Lasso. To ensure that the effects of the predictors
are comparable across variables we assume that each predictor has been
standardized by dividing by their standard deviation. This is not a
restriction when determining the importance of selected variables as
the internal structure of the predictors (and their pairwise
correlations) is preserved \citep{efron2004least}. Essentially, this
approach mimics the adaptive Lasso of \citet{Zou2006} and ensures that
the coefficients are equally penalized by the $\ell_1$ penalty in
\eqref{eq:lasso}. The variables can then be ordered according to their
absolute size such that $|\beta_{(1)}| \geq |\beta_{(2)}| \geq \cdots
\geq |\beta_{(p)}|$.

We are interested in testing hypotheses of the form
\begin{equation}
H_0 : \beta_{(k)} = 0,  \label{eq:null}
\end{equation}
where $\beta_{(k)}$ is the $k$th most important feature.  Note that
contrary to classical hypothesis testing we are not testing hypotheses
about \emph{specific} predictors but we are testing whether the $k$th
identified parameter is equal to zero.

As a test statistic for the $k$th selected parameter we use the
corresponding coefficient obtained from the Lasso, $|\beta_{(k)}|$.  In order to derive the distribution of the $k$th identified predictor
we permute response vector $\by$ to remove any associations to the set
of predictors while retaining the individual structure and
correlations among the set of predictors \citep{Manly2006}. This
randomization is undertaken a large number of times and for each
randomization we run the Lasso on the permuted data to obtain a
distribution of coefficients for the $k$th identified predictor. Note
that this approach might result in different variables being selected
for each permutation and this matches the null hypothesis
\eqref{eq:null} where the focus is on finding out whether the
$k$th identified predictor is significant.

The algorithm for testing the null hypothesis \eqref{eq:null} using a
randomization approach can be summarized as follows:
\begin{enumerate}
\item Start by scaling the predictors so their effects are comparable,
  i.e., 
$$X_j = X_j / \sqrt{\V (X_j)},$$
where $X_j$ is the $j$th column of the design matrix $X$.
\item Fit a Lasso regression model to the original data and extract
  the coefficients for the selected predictors, $\hat{\beta}_{(1)},
  \hat{\beta}_{(2)}, \ldots$.
\item Let $B$ be the (large) number of randomizations to perform to
  determine the distribution of the coefficients under the null
  hypothesis. For each $b\in B$ do the following:
\begin{enumerate}
\item Permute the response vector, $\by^b$, and fit a new Lasso
  model using $\by^b$ as the response.
\item Extract the coefficients for the predictors for this model, $\hat{\beta}_{(1)}^b,
  \hat{\beta}_{(2)}^b, \ldots$.
\end{enumerate}
\item For each feature, $k$, we can compare the size of the
  coefficient identified in the original dataset, $|\hat{\beta}_{(k)}|$
  to the distribution of the coefficients found for the $k$th feature
  for the permuted data, $|\hat{\beta}_{(k)}^1|, \ldots, 
  |\hat{\beta}_{(k)}^B|$. The $p$-value for the $k$th feature is then
  the fraction of coeffients under the null that are larger than or
  equal to $|\hat{\beta}_{(k)}|$:
\begin{equation}
p\mbox{-value} = \frac{1 + \sum_{b=1}^B \1_{  |\hat{\beta}_{(k)}^b|
    \geq  |\hat{\beta}_{(k)}| }}{B+1} \label{formula:pval}
\end{equation}
\end{enumerate}
Note that by construction the $p$-value defined by
\eqref{formula:pval} controls the error rate since we are simulating
the distribution on the null hypothesis.

\subsection*{Feature selection and collinearity}
In applications such as genetics, the number of predictors, $p$, is
often several orders of magnitude larger than $n$ and we are typically
more concerned with identifying \emph{some} associations rather than
finding \emph{all} associations since any relationships found will
often be the focus of subsequent specific experiments to verify the
genetic function. Thus, our interest is on hypothesis generation
rather than testing a specific hypothesis: is it possible to identity
one or a few potentially interesting or relevant predictors, in a
scenario with little or no knowledge of the individual predictors.

It is well-known that the Lasso has a weakness when dealing with
correlated predictors: it will only select one of a group of collinear
predictors to represent their effect on the response
\citep{zou2005regularization} and the coefficient for the selected
variable will have high variance \citep{meinshausen2008hierarchical}.
Collinearity poses a problem if we are intersted in making inferences
about specific predictors because the corresponding standard error
becomes unstable. However, in the present setup collinearity among
predictors means that while we may not be able to identify all
predictors that are associated with the outcome, the size of the
coefficients for the variable that are selected essentially represent
the maximum effect size of the cluster of correlated predictors.

In particular, if $p\gg n$ then we may be interested in just finding
\emph{one} association between a predictor and the response in which
case our primary focus is on the null hypothesis
\begin{equation}
H_0 : \beta_{(1)} = 0. 
\label{eq:null1}
\end{equation}

\subsection*{Choice of penalty parameter $\lambda$}
The Lasso involves a penalty parameter, $\lambda$, that determines the
amount of shrinkage that is applied to the parameters. In our
approach, we recommend using the same procedure for estimating
$\lambda$, that the reseacher would typically use together with a
Lasso regularization model. Typically, $\lambda$ is determined through
some form of $K$-fold cross-validation where $\lambda$ is chosen such that it
minimizes the cross-validation error.

\begin{equation}
CV(\lambda) = \frac1K \sum_{i=1}^N |y_i - \hat{f}^{-\kappa(i)}(\bx_i,\lambda)|,
\end{equation}
where $\bx_i$ is the set of covariates for individual $i$,
$\hat{f}^{-\kappa(i)}$ is the expected value from the Lasso regression
based on the folds/parts of the data that does not contain observation $i$.

There are two considerations to consider here: First, we know that we
are in a situation where the majority of the variables are unrelated
to the outcome. Thus we want to ensure, that the majority (or possibly
all) of the variables are shrunk to zero. This is particularly true
under the null hypothesis where there is no association between the
variables and the outcome due to randomization. Secondly, we want to
ensure that when there is indeed an association between some of the
variables and the outcome then we want $\lambda$ to be small enough
that the effect of the associated variables are not shrunk too much.

In the following we have worked with a 10-fold cross-validation using
the mean absolute error as loss function since our analyses have shown
that this gives good stable results. Using the mean absolute error
instead of the traditional mean squared error makes the
cross-validations less sensitive to spurious extreme fits which is
often the case under the null, where none of the variables are
associated with the outcome.  However, other approaches would be
directly applicable as well at the researchers discretion.

One possible approach here, that would substantially reduce the
computation time would be to use the same $\lambda$ for the
randomization samples as was used in the original dataset. That
ensures that we only have to do cross-validation once (for the
original data) and not for each randomization. Also, the estimate of
$\lambda$ under the null hypohtesis (where the randomization approach
assumes that \emph{none} of the predictors are associated with the
outcome) can be rather unstable since there really is no obvious
minimum for the cross validation procedure to reach.

\subsection*{Including external predictors}
In some situations it is of interest to use the Lasso to select
predictors from among a subset of the predictors. For example, some
predictors might be related to the design of the experiment (and we
wish to force them to be part of the modeling) while the remaining
predictors contains the variables we wish to identify/select from
because we have no prior knowledge about them.

Thus, we consider two sets of predictors, $\tilde{X}$ and $X'$ such
that $X = [\tilde{X}|X']$ where $\tilde{X}$ are the predictors that we
wish to force into the model while $X'$ are the predictors from which
we wish to make a selection. If we apply the Lasso to $X$ we will
select variables from the full set of predictors which means that some
of the potential confounders that we wish to include are disregarded
in the modeling.

There are two obvious alternatives to just using the classical Lasso
in this situation. One is to make a two-step procedure where we first
fit a model using only the predictors in $\tilde{X}$, i.e., 
\begin{equation}
\by = \tilde{X}\tilde{\bbeta} + \bepsilon. 
\label{eq:stepprocedure}
\end{equation}
From \eqref{eq:stepprocedure} we extract the residuals, $\bm{r} = \by
- \tilde{X}\hat{\tilde{\bbeta}}$, and use the residuals as outcomes in
combination with the procedure described above. This approach ensures
that the (marginal) effects of the predictors in $\tilde{X}$ are
removed and and features that are subsequently found are conditional on
these marginal effects. The disadvantage is that the effects of the
predictors in $\tilde{X}$ and $X'$ are assumed to be working
independently of each other, which may not be realistic.

The adaptive (or weighted) Lasso of \citet{Zou2006}  uses a weighted
penalty resembling \eqref{eq:lasso}
\begin{equation}
\hat{\bbeta}=\underset{\bbeta \in \mathbb{R}^p}{\operatorname{arg\,min}}
\left\{ \frac{1}{2}\|\by-X\bbeta\|^{2}_{2}+\lambda\|\bw\bbeta\|_{1} \right\},
\label{eq:adaptivelasso}
\end{equation}
where $\bw$ is a vector of non-negative weights for each predictor.
Typically, the individual weights are set to $w_j =
1/|\hat{\beta_j'}|^\nu$, where $\hat{\beta_j'}$ is the univariate
regression coefficient estimate of parameter $j$ and $\nu> 0$. If a
weight is set to zero then the corresponding predictor will not be
penalized. Thus if the weights of the parameters in $\tilde{X}$ are set
to zero then the corresponding variables enter the model without being
shrunk towards zero. 

\subsection*{Family-wise error rate for multiple inference results}
The $p$-value obtained from \eqref{formula:pval} controls the error
rate by construction as mentioned above. However, the error rate
control is on a per-hypothesis basis so if we are only aiming on
making inference for the ``best'' selected predictor --- corresponding
to the hypothesis \eqref{eq:null1} --- then the proposed procedure
yield the correct error level.

However, in some situations it is of interest to extract inference on
as many predictors as possible and simultaneously control the
family-wise error rate. We can still use \eqref{formula:pval} to
obtain individual $p$-values for the first feature, the second
feature, etc., and Holm's step-down procedure can control the
family-wise error rate in those situations \citep{Holm1979}. In its
original form, Holm's procedure orders the $p$-values and makes
sequential comparisons from the smallest $p$-value to the largest
until the first occurrence of a null hypothesis that fails to be
rejected. All other subsequent hypotheses are also not rejected.

Here we suggest a slightly different version of Holm's procedure. The
$p$-value of the $k$'th selected feature is compared against
$$\frac{\alpha}{m+1-k},$$
where $\alpha$ is the overall significance level desired and $m$ is
the number of hypotheses tested. The significance level for each
feature is identical to the level used by Holm's procedure but unlike
that we do not order our $p$-values according to size but test them in
the order that they are selected by our procedure.  This ensures that
we still control the family-wise error rate (by the exact same
arguments as used by \citet{Holm1979}) since it becomes more difficult
to reject hypotheses by \emph{not} ordering them according to the size
of the $p$-value. Also, since we keep the order of the features it
prevents us from ending with a result where, say, the first and third
selected features are significant but the second selected feature is
not.

%\end{methods}

\section{Results}

\subsection*{Simulation studies} \label{sec:results}

To examine the validity of our procedure, we performed simulation
studies to assess the performance of the sensitivity and
specificity. For each setting, we simulated 100 data sets with only $50$
observations generated under the linear model,
$$
\by = X\bbeta + \bepsilon, 
$$
where $\bepsilon \sim N(0, \bI)$ and where $X_{ij} \sim
N(0,1), \; i=1,\ldots, 50, \; j=1, \ldots, p$.  The unit variance is
not really a restriction in this case as we start the analysis
algorithm by standardizing each predictor (as outlined in the methods
section above).  The number of predictors were varied from 1000 to
250000 to represent various datasets.  

Initially we assume that there is just a single predictor that is
associated with the response and we vary the corresponding regression
coefficient, $\beta$, from 0 to 1.5. This corresponds to a partial
correlation coefficient between the predictor and the outcome ranging
from 0 to 0.83. With these data we follow the procedure described
above where each combination of predictors, $p$, and associated
coefficient, $\beta$, is run 100 times and for each combination we use
100 permutations to determine the power of the feature identified as
the most important corresponding to the hypothesis $\beta_{(1)} = 0$.
In all computations a significance level of $\alpha=0.05$ was used.
In order to see how collinearity influences the power we run
simulations with the same association between a single predictor and
the outcome but we let that predictor be part of a cluster of ten
correlated predictors such that $\mbox{cor}(X_i, X_{i'}) = \rho$.
Here, $\rho=0.5$ or $\rho=0.95$ to represent moderately and highly
correlated predictors.

\begin{figure}[tbp]
   \centering
   \includegraphics[width=.95\columnwidth]{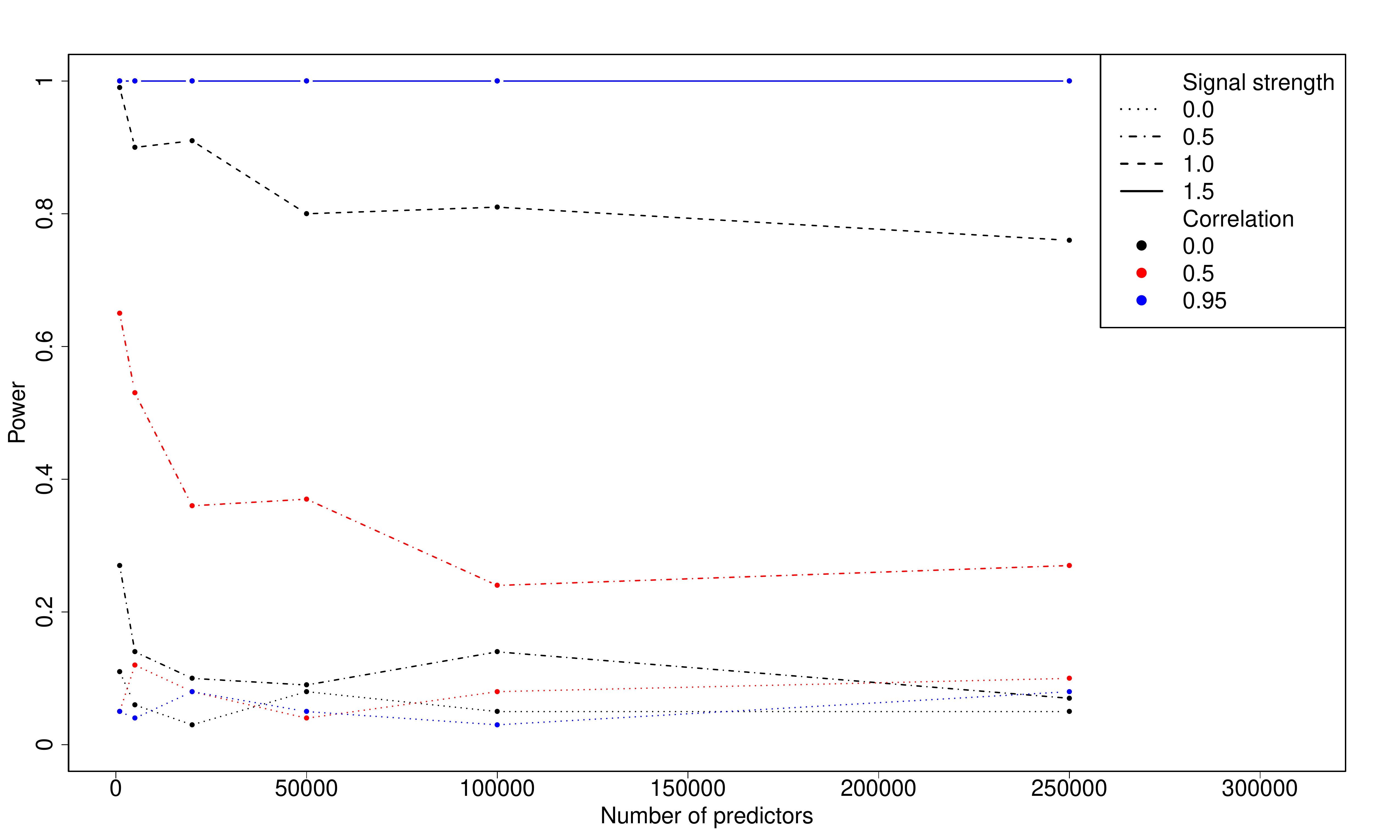} 
   \caption{Power to detect the first identified predictor selected by
     the Lasso for varying number of predictors and varying effect
     size of the predictor for a dataset containing 50
     observations. Each point is the average from 100 simulations,
     each using 100 randomizatons to compute the $p$-value. The black
     lines correspond to completely independent predictors, the red
     lines are when there is some collinearity between a group of
     predictors while the blue lines represent the situation with a
     group of highly correlated predictors.}
   \label{fig:pvals}
\end{figure}

We first focus on two primary abilities of the model: the power to
identify the first selected predictor and the impact of correlated
predictors in the model. The simulation results are shown in Figure
\ref{fig:pvals}. The simulations show three clear results: 1) for
moderate to high effect size (i.e., when $\beta$ is 1--1.5) we obtain a
high power even when the number of predictors is large, 2) when the
causal gene is part of a correlated cluster of predictors then the
power is higher than if the gene is independent, and 3) generally
there is a fall in power as the number of predictors increase although
this drop appears to stabilize and level off as the set of predictors
increases.

The power is 100\% for an effect size of 1 or 1.5 (the lines overlay each
other in Figure~\ref{fig:pvals}) except for the independent
predictors. That means that even when there is a ratio of predictors
to observations, $p/n$, of 5000 then we are still able to identify a
signal in the data even with just 50 observations. An effect size of 1
corresponds to an effect of 1 standard deviation so we are even able
to identify the presence of a gene with realistic biogical effects. 

For correlated data we see that the power is \emph{larger} than the
power observed from independent predictors. This is caused by the fact
that the causal predictor is part of a cluster; with correlated
predictors, then each of the predictors in the cluster has a
probability of showing an improved association to the outcome just by
chance. Thus, we essentially improve our power to detect \emph{one} of
the predictors in the cluster at the cost of perhaps not detecting the
true one but at least one that is highly correlated to it. If the
causal predictor was a singleton and not part of a cluster of
correlated predictors then the power results resemble the results from
the uncorrelated data (results not shown).

Initially the power drops for all situations as the number of
predictors increase (making it harder to identify that there is in
fact a true association among one of the predictors and the outcome)
but the overall power seems fairly constant once the number of
predictors reach around 50000 (of which just one is associated with
the outcome). Not surprisingly, this drop is more noticeable when the
true association has a moderate effect than when the effect is smaller
(where the power is consistently low) or larger (where the power is
consistently high).  Figure \ref{fig:pvals} also shows that we are
generally able to control the family-wise error rate at the desired
significance level (0.05). Perhaps most surprisingly is the consistent
high power for moderate to high effect sizes ($\beta=1$ or
$\beta=1.5$) considering this is based on just 50 observations.

Figure~\ref{fig:pvals} focused on the evidence of the first/most
important predictor. Once that has been identified we would like to
determine the power to detect the second most important predictor and
its power. In our simulations we assumed that there was always
one causal predictor with a corresponding parameter of $\beta_1 = 1.5$
and we varied the effect of another predictor, $\beta_2$, with values
ranging from 0 to 1.5 as previously. Following the setup above we
also allow for correlations and assume that the two predictors are
actually part of the same cluster. 

\begin{figure}[htbp]
   \centering
   \includegraphics[width=.95\columnwidth]{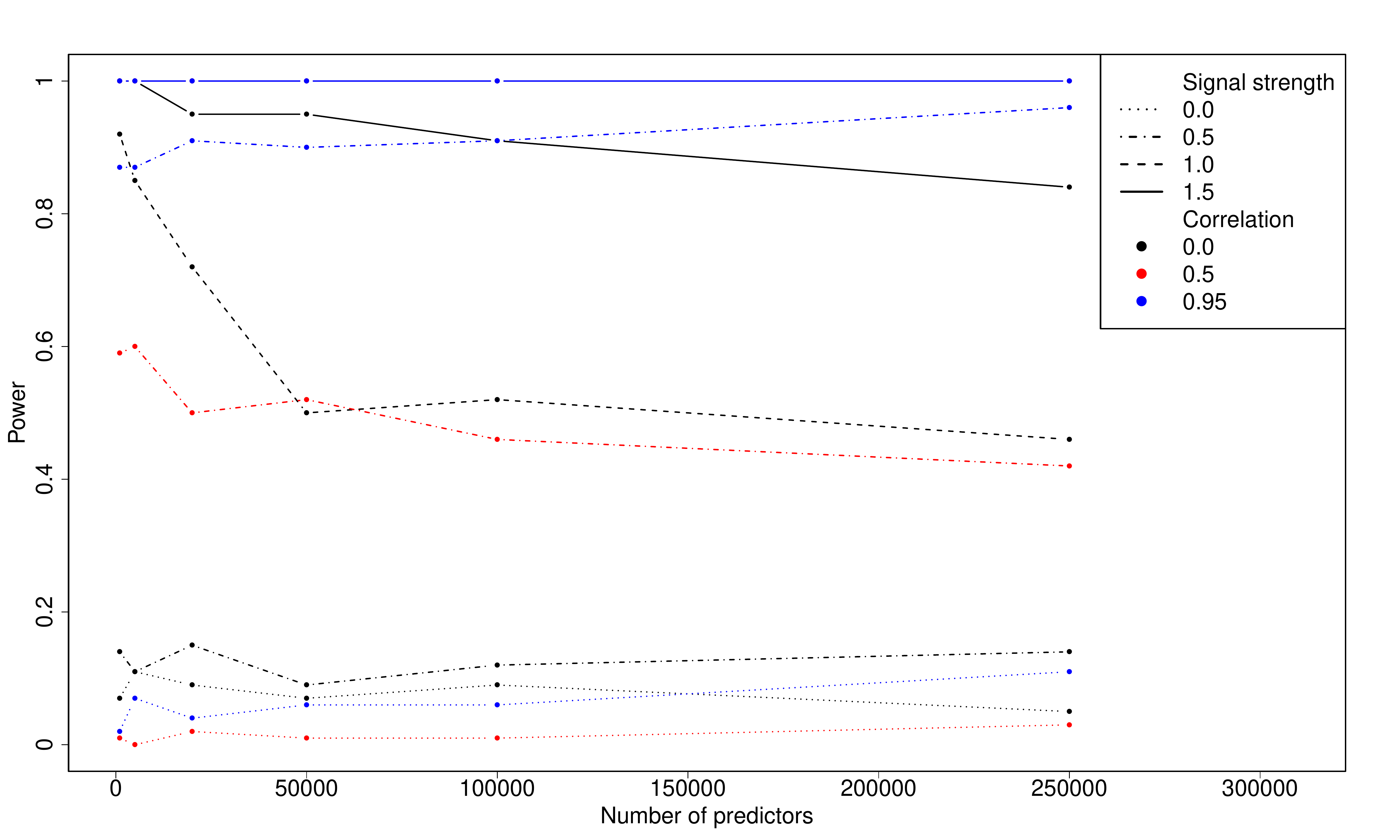} 
   \caption{Power to detect the \emph{second} identified predictor
     selected by the Lasso for varying number of predictors and
     varying effect size of the predictor for a dataset containing 50
     observations. The dataset contains two predictors from the same
     group/cluster that are associated with the response, one of which
     is fixed with an effect size of 1.5.  Each point is the average
     from 100 simulations, each using 100 randomizatons to compute the
     $p$-value. The black lines correspond to completely independent
     predictors, the red lines are when there is some collinearity
     between a group of predictors while the blue lines represent the
     situation with a group of highly correlated predictors.}
   \label{fig:2ndpvals}
\end{figure}

Figure~\ref{fig:2ndpvals} shows the power to detect the \emph{second
  most important} variable when there are two potential causal
variables from the same cluster. Overall, we see the same trends as we
saw for the most important variable in Figure~\ref{fig:pvals} except
that the power is generally lower. 
Looking at the solid line corresponding to the largest effect of the
second most important variable we see a marked decline in power as the
number of predictors increase if the predictors are all
independent. The decline is not very surprising as both of the two
causal predictors have the same effect size (both $1.5$) so the second
most important variable will by design be the smallest of those two
(after sampling error has been added). Thus, due to random variation,
the estimated effect of the second is likely to be slightly smaller
than the true effect (ie., biased downward) and contrary the effect of
the most important variable will be slightly biased upward. However,
the drop in power for the second most important variable is still
substantial even for high effect sizes except when the predictors are
correlated.  Figure~\ref{fig:2ndpvals} also shows that the error rate
is controlled around the 5\% level for independent predictors when
there is only one predictor that is truly associated to the response
(black dotted line). As for the results from the first identified
predictor, Figure~\ref{fig:pvals}, we get that correlated predictors
increase the power to identify subsequent predictors.

\begin{figure}[tbp]
   \centering
   \includegraphics[width=.95\columnwidth]{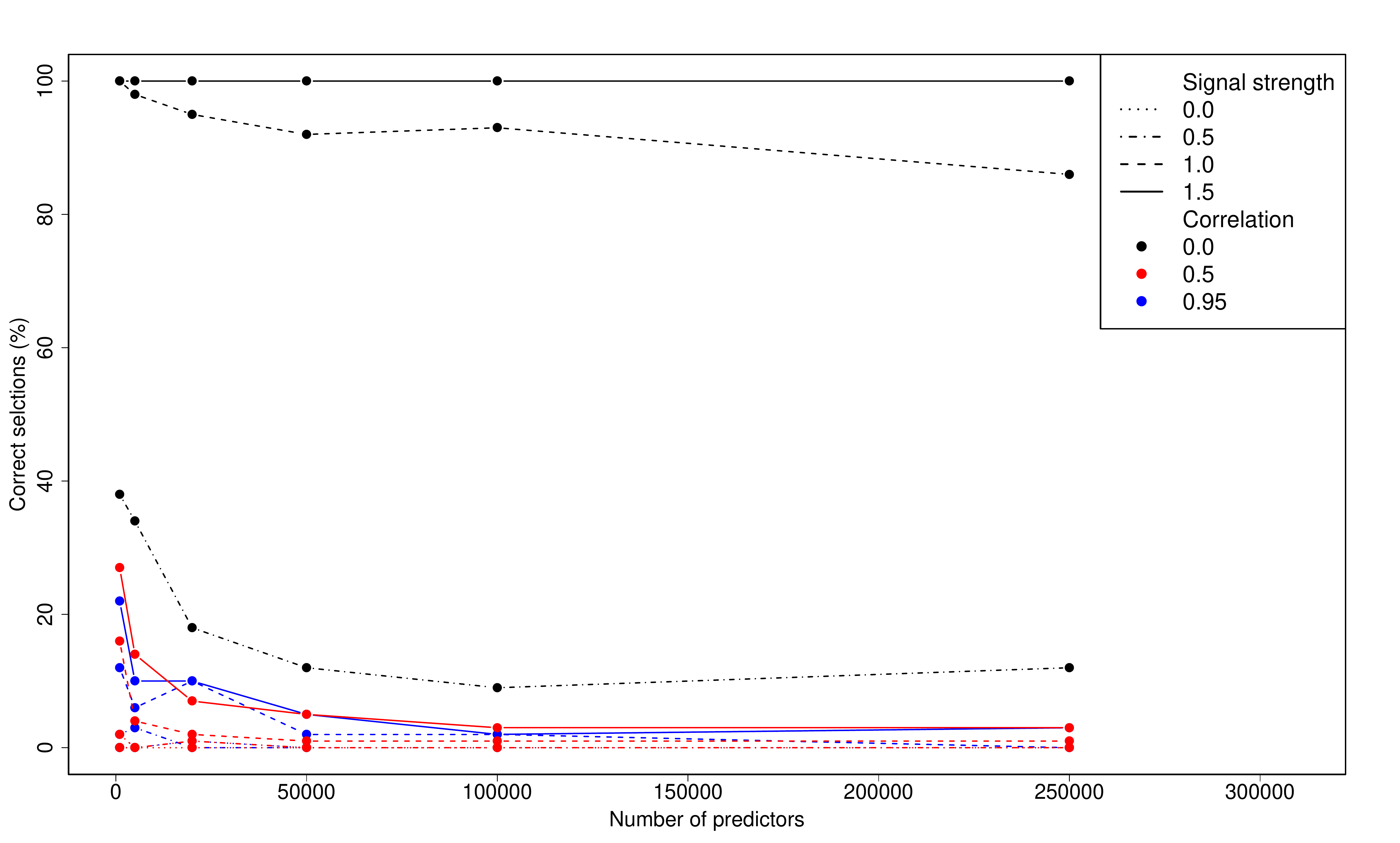} 
   \caption{Feature identification precision for the Lasso to identify
     the single true underlying (most significant) predictor for
     varying number of predictors and effect sizes. Each point is the
     average from 100 simulations. The black lines correspond to completely
     independent predictors, the red lines are when there is some
     collinearity between a group of predictors while the
     blue lines represent the situation with a group of highly
     correlated predictors.}
   \label{fig:detecttrue}
\end{figure}

As highlighted in the methods section, the proposed procedure computes
the evidence for the identified features but does not directly provide
significance inference for a particular predictor. In practice,
however, it is also of interest how well the Lasso approach identifies
the predictors, that are truly associated with the response. While
this has been investigated in several papers
\citep{efron2004least,hastie2005elements,
  hesterberg2008least,meinshausen2009p}, we include results using the
same simulation setup as described above for direct comparison with
the simulation results shown in Figure \ref{fig:pvals}. The results
are shown in Figure \ref{fig:detecttrue} which clearly shows that the
precision increases with increasing signal strength and decreases when
the true predictor is part of a cluster of predictors. In the latter
case, it is generally one of the the predictors from the correlated
cluster that is identified, so while we may not identify the specific
variable we do identify the cluster (results not shown).

\subsection*{Application to prostate cancer data}

We apply our proposed method to the prostate cancer example from
\citet{Lockhart:2013fk}, a subset  of  67 samples  and 8
predictors from a larger dataset. The outcome is the logarithm of a prostate-specific
antigen and we want to determine the association between the outcome
and any predictors. Data can be found at
\verb+www-stat.stanford.edu/~tibs/ElemStatLearn/+. 
The setup for this dataset is simpler than the situation
mentioned in methods section since $p<n$ so it is possible to analyze
all predictors simultaneously by using, say, a classical multiple
regression model. We compare the results from our proposed method with
three other analyses approaches: simple multiple regression, using the
Lasso in combination with the covariance test statistic of
\citet{Lockhart:2013fk} and the multi-split method of
\citet{meinshausen2009p}. The multi-split method essentially consists
of variant of 2-fold cross-validation: First, Lasso regression is
applied to a part of the dataset to select a list of predictors, and
secondly an ordinary multiple regression (using the selected
predictors) is applied to the remaining data to obtain $p$-values for
the selected predictors.

Note that unlike our proposed approach, both the multiple regression,
the Lasso-covariance test, and the multi-split method are all designed
to test hypotheses about each specific predictor. However, the multi-split
method is vulnerable to the selection obtained from the first split
which gives rise to rather substantial variance of the $p$-values in
this dataset. Splitting the sample repeatedly and getting a set of
$p$-values has also been suggested by the authors, but ``it is not
obvious, though, how to combine and aggregate the results.''
\citep{meinshausen2009p}. Here, we have just averaged the $p$-values
obtained for each of the sets (with non-selected predictors getting a
$p$-value of 1).

\begin{table*}[t]
\centering
\caption{Table of $p$-values for each predictor obtained from simple linear
  regression (LR), multiple
  regression (OLS), the multi-split method (SPLIT), the covariance test method (COV)
  and the proposed method (RAND). Last column shows the marginal
  Pearson correlation between each predictor and the outcome. 
Predictors are listed in order they are selected by the 
randomization procedure.}
\label{tab:pval}
\begin{tabular}{rrrrrrc}
    \hline
Predictor  &LR $p$ & OLS $p^\ast$ & COV $p$ & SPLIT $p$ & RAND $p$ & Correlation with outcome \\ 
  \hline
%(Intercept) & 2.42 & 0.00 & 0.00 & 0.00 & 1.00 \\ 
  lcavol &0.00 & 0.00 &  0.00 & 0.00 %& 0.00 
  & 0.00& 0.73 \\ 
  svi & 0.00 & 0.05& 0.17 & 0.29 %& 0.29 
& 0.34 & 0.56  \\ 
  lweight & 0.00 & 0.00 &0.05 & 0.06 %& 0.10 
& 0.14& 0.49 \\ 
  lcp & 0.00 &  0.09&0.05 & 0.89 %& 0.25 
& 0.07 & 0.49\\ 
  pgg45 & 0.00  &0.24&  0.35 & 0.86 %& 0.26 
& 0.01& 0.45 \\ 
  lbph & 0.03 & 0.05&0.92 & 0.51 %& 0.37 
&  0.01& 0.26\\   
 age & 0.06 & 0.14& 0.65 & 0.92 %& 0.22 
& 0.04 & 0.23 \\  
  gleason & 0.00 & 0.88& 0.97 & 0.97 %& 0.12 
& 0.58& 0.34 \\ 
   \hline
\multicolumn{6}{l}{$^\ast$ $p$-values for OLS are shown after backward elimination.}
\end{tabular}
\end{table*}

Table~\ref{tab:pval} shows that the three methods that accommodate
multiple predictors and test hypotheses about each specific predictor
(OLS, multi-split and the covariance test) generally give the same
results and all identify the same predictor, \emph{lcavol}, as being the most
important. The same result is obtained with our proposed method, where
the first selected variable is highly significant (and turns out to be
\emph{lcavol}). The simple marginal regression analyses show that
virtually all predictors are associated with the response, but the
other approaches reduce the number of significant variables partly due
to correlation among the predictors. Our proposed randomization test
identifies slightly more variables than the covariance test (and a few
more than the multi-split procedure) and seems to lie somewhere
between the ordinary multiple regression model and the covariance test.

It is only the covariance test that places any real emphasis on the
\emph{lcp} predictor in part because of a high degree of collinearity
to \emph{lcavol} (the correlation coefficient between \emph{lcavol}
and \emph{lcp} is $0.692$, which is the second largest among the
predictors).  The largest pairwise correlation is between \emph{pgg45}
and \emph{gleason} and is 0.757 but the association to the outcome is
less for those two variables so the impact of their collinearity on
the $p$-values is less noticeable.

 \section{Discussion}
 We have proposed an algorithm to aid in making inference and
 subsequently relevant hypotheses concerning data with (many) more
 predictors than samples. It greatly extends the results from
 shrinkage regression methods to be able to assign a $p$-value to
 findings from regularized regression methods that combine variable
 selection and estimation.

 When the number of predictors $p$ is larger than then sample size
 $n$, regularized regression methods can be used to identify a sparse
 model and to provide stable parameter estimates. The standard Lasso
 suffers from several problems in this regard, in particular that it
 is not consistent in cariable selection and that the limiting
 distribution of the estimates are non-standard and cannot be directly
 derived. To overcome these shortcomings \citet{fan:li:2001} and
 \citet{Zou2006} have presented versions of regularized regression
 that not only have asymptotic oracle properties but also have
 consistency in variable selection and asymptotic normality of the
 estimators.  However, inference based on these improved
 regularization methods for \emph{finite samples} generally perform
 poorly even when the effects are large \citep{minn:etal:2011}.

 An obvious difference between the $p$-values obtained from ordinary
 least squares or from other regularized regression inference
 approaches is that we pursue investigating a null hypothesis that is
 not concerned with a specific (set of) predictor(s) but is concerned
 solely with evaluating the evidence that the results obtained from
 regularized regression are stronger than what would be expected if
 none of the predictors were associated with the response.  Classical
 OLS-style $p$-values are extremely useful but it can be argued that
 for the majority of problems in genomics --- where there is often
 little or no prior information about \emph{any} of the possible
 predictors --- the focus is on hypothesis generation and variable
 selection and not on testing specific hypotheses. Hence our focus
 (and the proposed procedure) is on hypothesis generation and not on
 inferences about specific variables.

 Our simulations show that --- even with just 50 observations --- the
 proposed procedure has substantial power to identify at least one
 associated predictor among a set of 50--250 thousand predictors
 without serious performance decline (Figure \ref{fig:pvals}). This
 suggests that not only can we can identify relevant predictors within
 a large set of irrelevant predictors --- we can also attach a level
 of significance so we can evaluate whether our findings are indeed
 likely to be true (and relevant) predictors, that are associated to
 the outcome. Our results extend to the situation where there are
 multiple causal variables (Figure~\ref{fig:2ndpvals}) although the
 power declines slightly more rapidly with the number of predictors
 when testing the importance of subsequent features. The results
 regarding the second most important variable shown in
 Figure~\ref{fig:2ndpvals} was based on two causal variables in the
 same cluster. If we run the same analyzes where the two causal
 variables are in two different clusters we get essentially the same
 results as shown for the uncorrelated data in
 Figure~\ref{fig:2ndpvals} (results not shown).

 We have run the same simulations with a reduced dataset of 30
 observations instead of 50 observations and while the power obviously
 is lower we see the same overall trends as shown above. Thus, even
 with fairly small datasets we have decent power to identify a
 variable that truly is associated to the outcome.

 It can be argued that for genetic data where many genes may follow a
 similar pattern of expression and thus be correlated, Lasso is not an
 obvious choice for variable selection. If the purpose of the
 selection is indeed to identify a whole cluster of co-expressed
 genes, other shrinkage algorithms could be better suited. However if
 the genes selected by the Lasso are seen as the most influential
 representative of a cluster there are numerous ways to identify
 others in the same cluster. Since the whole nature of our approach
 is not to identify specific variables this is not a problem \emph{per
 se}. In fact, Figures~\ref{fig:pvals} and \ref{fig:2ndpvals} show
that the power to detect \emph{something} from a clustered set of
predictors rise dramatically so clustered predictors will just increase our
power to identify at least one of the predictors from the cluster.

When comparing methods applied to the prostate dataset, it was clear
that all four methods (the proposed approach, ordinary least squares,
the covariance test statistic, and the split method) are able to
identify the variable lcavol as among the most important and a small
$p$-value is assigned to this predictor. Note that while the four
methods differ somewhat in their results they also test different
versions of the null hypothesis. The split method consistenly assigns
a small $p$-value to the lcavol predictor but the remaining seven
predictors obtain $p$-values estimated with very large uncertainties.
The randomization approach generally yields lowest $p$-values which is
partly due to the fact that the variables are indeed mostly correlated
with the outcome, and partly due to that the null hypothesis is
different (in particular it assumes that \emph{none} of the predictors
are associated with the outcome).

We believe the proposed method could also be applied in for instance
Genome-Wide Association Studies (GWAS), where Lasso is already widely
used, but significance testing typically done outside the Lasso
domain, as done by \cite{Wu2009}.

\section{Conclusion}
We have presented a method that can be used to make inference about
variable selection results from regularized regression models such as
the Lasso. We show that it has high power to infer evidence that the
selected features are not chance findings --- even when the number of
predictors is several orders of magnitude larger than the number of
observations, i.e., when $p>n$. The method controls the family-wise
error rate and can essentially be used with any regularization
method. The proposed method is relevant for situations where
regularized regression methods is used as part of the statistical
modeling to identify features for subsequent analyses.

%\bibliographystyle{natbib}
%\bibliography{paperref}

\end{document}